\documentclass{IEEEoj}
\usepackage{cite}
\usepackage{amsmath,amssymb,amsfonts}
\usepackage{graphicx,color}
\usepackage{textcomp}
\usepackage{booktabs}
\usepackage{float}
\usepackage{array}
\usepackage{url}
\def\OJlogo{}
\def\BibTeX{{\rm B\kern-.05em{\sc i\kern-.025em b}\kern-.08em
    T\kern-.1667em\lower.7ex\hbox{E}\kern-.125emX}}

\begin{document}
\receiveddate{XX Month, XXXX}
\reviseddate{XX Month, XXXX}
\accepteddate{XX Month, XXXX}
\publisheddate{XX Month, XXXX}
\currentdate{XX Month, XXXX}
\doiinfo{OJCS.2026.XXXXXXX}

\title{Measuring Tabular Data Formats Cold: A Replicated Cold-Cache Benchmark of Serialization, Scan, and Index Performance on Real Datasets}

\author{ASRA FATHIMA\IEEEauthorrefmark{1} AND KRISTOFFER ROBERTS\IEEEauthorrefmark{1}}
\affil{University of the Cumberlands, Williamsburg, KY 40769 USA}
\corresp{CORRESPONDING AUTHOR: Asra Fathima (e-mail: afathima39904@ucumberlands.edu).}
\authornote{This research received no external funding. This work has been submitted to the IEEE for possible publication. Copyright may be transferred without notice, after which this version may no longer be accessible.}
\markboth{Measuring Tabular Data Formats Cold}{Fathima and Roberts: Cold-Cache Benchmark of Tabular Data Formats}

\begin{abstract}
Choosing a file format for tabular data is a routine engineering decision, yet published comparisons often time reads served from the operating-system page cache, use one session, or rely on a single synthetic dataset. This article benchmarks fifteen format and access-path configurations (CSV, JSON Lines, Pickle, Feather, Parquet, ORC, DuckDB, and SQLite, with several access paths) on two real public datasets, New York City taxi trips and United States airline on-time records, at 50,000 to 3 million rows. Every measurement uses an explicitly cold cache, randomized execution order, and three independent sessions (with restarts on the laptop), and is analyzed with bootstrap intervals and Holm-adjusted permutation tests. Parquet answered a filtered aggregation 2.6 to 11 times faster than CSV, while JSON Lines was 10 to 27 times slower. A secondary index made SQLite queries 2.6 to 4.8 times slower on randomly ordered data and helped only on data sorted by the filtered column, at low selectivity. A row filter inside the Parquet reader was significantly faster than a projected scan on flight data from one million rows and on both datasets at three million, and row-group skipping needs several row groups, which the default writer produced only at 3 million rows. A 4~GB Linux host ran out of memory on configurations that a 24~GB laptop completed and reversed some orderings, such as Pickle versus Parquet. Code, derived data, and raw timings are released.
\end{abstract}


\maketitle

\section{INTRODUCTION}
\IEEEPARstart{D}{ata} engineers and analysts choose among many file formats for tabular data: delimited text, JSON Lines, language-native serialization such as Python's Pickle, Arrow-based Feather, columnar files such as Parquet and ORC, and embedded databases such as SQLite and DuckDB. Practitioner advice is plentiful, but the evidence behind it is uneven. Three weaknesses recur in format comparisons. First, reads are frequently timed with the file already in the operating-system page cache, so the numbers describe memory bandwidth rather than storage. Second, each configuration is measured once or in a single run, which cannot separate a real difference from drift in machine state. Third, the data is synthetic or comes from one source, so conclusions may not transfer to real columns with real value distributions.

This article addresses these weaknesses directly. We measure how long it takes to obtain the answer to a filtered aggregation, from an on-disk file whose cached pages have been evicted, for fifteen format and access-path configurations. The experiments use two real public datasets, repeat every measurement in three independent sessions separated by restarts of the primary laptop, and report ratios with confidence intervals and Holm-adjusted permutation tests. We add a selectivity sweep, data sorted by the filter columns, and a second host with only 4~GB of memory.

We organize the study around four research questions.
\begin{itemize}
\item \textbf{RQ1.} Which formats answer a filtered aggregation fastest from a cold cache, and how do the differences change from 50,000 to 3 million rows?
\item \textbf{RQ2.} How much of Parquet's advantage comes from reading only the needed columns, and how much from skipping row groups?
\item \textbf{RQ3.} When does a secondary index in SQLite help, as a function of data order and selectivity?
\item \textbf{RQ4.} How stable are the conclusions across sessions, datasets, and a small-memory host?
\end{itemize}

The contributions are as follows. (1) A cold-cache measurement protocol for a laptop and a small Linux host, with self-tests that verify the cache was actually cold, and with a randomized, multi-session design. (2) Results on two real datasets with interval estimates and Holm-adjusted permutation tests, including the ratio ranges for every pairwise comparison we draw conclusions from. (3) Two findings that contradict common assumptions: a secondary index slowed SQLite queries at every selectivity we tested when rows were in random order, and Parquet row-group skipping was unavailable in default-layout files below 3 million rows because they held a single row group, while a row filter inside the reader was significantly faster than a projected scan on flight data from one million rows and on both datasets at three million. (4) Evidence that conclusions about pairwise orderings do not always carry across hosts. (5) Public release of code, derivation scripts, and raw timings.

Section~II reviews related work, Section~III describes the method, Section~IV reports results, Section~V discusses practical guidance, Section~VI lists limitations, and Section~VII concludes.

\section{BACKGROUND AND RELATED WORK}
\subsection{Format comparisons}
Zeng et al.\ evaluated Parquet and ORC in detail, examining encodings, compression, and the effect of file layout parameters on decoding speed and scans on modern hardware~\cite{zeng2023}. Their emphasis is the internal design of the two columnar formats, on synthetic and real data. Plase et al.\ compared Avro and Parquet for storage in HDFS~\cite{plase2017}. The earlier column-store literature established why scanning only the needed columns pays off in analytical queries~\cite{abadi2008}. Our study complements these works by comparing across format families, from text through binary row serialization to columnar files and embedded databases, using the same query and a cold cache, and by examining the access path as well as the file format.

DuckDB is an embeddable analytical database that reads Parquet directly and executes a vectorized engine in process~\cite{raasveldt2019}. SQLite is an embedded row-oriented relational database. Both appear here as query engines against which file-based pipelines in pandas~\cite{mckinney2010} can be compared.

\subsection{Rigorous performance measurement}
The performance-evaluation literature documents how easily measurements mislead. Georges et al.\ argued for statistically rigorous evaluation with repeated invocations and confidence intervals~\cite{georges2007}. Mytkowicz et al.\ showed that innocuous setup choices can bias results~\cite{mytkowicz2009}. Kalibera and Jones described how to allocate repetitions across levels, such as sessions and iterations within a session, so that variance at each level is accounted for~\cite{kalibera2013}. Hoefler and Belli proposed rules for reporting performance, including intervals around summary statistics and full documentation of the experimental setup~\cite{hoefler2015}. Maricq et al.\ showed that performance on nominally identical machines varies enough to change conclusions and proposed confidence-based methods for choosing repetition counts~\cite{maricq2018}. Hoenig and Heisey explained why post hoc power calculated from an observed result adds no information beyond the p-value~\cite{hoenig2001}. We follow these recommendations by randomizing execution order, using several sessions, reporting ratios with intervals, and avoiding post hoc power.

\section{METHOD}
\subsection{Configurations}
Table~\ref{tab:formats} lists the fifteen configurations. A configuration is a file format together with an access path. All are driven from Python with pandas 3.1 and PyArrow 26 on the main host, except DuckDB and SQLite, which execute the query in their own engines. We use the term \emph{projected scan} for reading only the columns the query needs, and \emph{filtered scan} for projected scan plus row filtering inside the Parquet reader. Selectivity is the fraction of rows that match the filter.

\begin{table}[t]
\centering
\caption{Format and access-path configurations. ``Cols'' = which columns are decoded, ``Filter'' = where rows are filtered, ``Agg.'' = where the group-by sum runs}
\label{tab:formats}
\small
\setlength{\tabcolsep}{2.2pt}
\begin{tabular}{@{}lccc@{}}
\toprule
Configuration & Cols & Filter & Agg.\\
\midrule
CSV, JSON Lines, CSV+gzip & all & pandas & pandas\\
CSV (usecols) & needed & pandas & pandas\\
Pickle, Feather, Parquet, ORC & all & pandas & pandas\\
Feather (projected), Parquet (projected) & needed & pandas & pandas\\
Parquet (filtered; also 100k groups) & needed & reader & pandas\\
DuckDB & needed & engine & engine\\
SQLite (no index) & needed & engine & engine\\
SQLite (indexed+ANALYZE) & needed & engine, index & engine\\
\bottomrule
\end{tabular}
\end{table}

\subsection{Datasets, query, and layouts}
We use two real public datasets. The first is the New York City Taxi and Limousine Commission yellow-taxi trip records for June 2026 (3,837,248 trips)~\cite{tlc}. The second is the United States Bureau of Transportation Statistics airline on-time performance data for January through July 2026 (4,120,343 flights, 43 columns)~\cite{bts}. The taxi file is the unmodified monthly Parquet file from the publisher; the flight data are the seven monthly CSV extracts (all 43 selected fields) concatenated, with the date column converted to a datetime type and no rows removed. For each size, rows are drawn without replacement with a fixed seed (42). The samples preserve the filter selectivity of the full file: the taxi query matches 1.05\% of the full month and 1.09, 1.07, 1.05 and 1.05\% of the 50,000, 250,000, 1,000,000 and 3,000,000-row samples; the flight query matches 0.88\% of the full data and 0.83, 0.87, 0.89 and 0.88\% of the samples. A synthetic dataset generated with a fixed seed is kept only as a controlled comparison and is never described as real.

Each dataset has one query: a conjunctive filter on two columns followed by a group-by sum. On the taxi data the filter is \texttt{payment\_type = 2 AND PULocationID = 132} (1.05\% of rows), grouped by vendor, summing the fare. On the flight data it is \texttt{OP\_UNIQUE\_CARRIER = 'UA' AND DEST = 'SFO'} (0.88\% of rows), grouped by day of week, summing arrival delay. The same answer is required from all formats, and a checksum of the answer is recorded for every run; mismatches would be flagged, and none occurred.

Sizes are 50,000, 250,000, 1,000,000 and 3,000,000 rows. Each real-data sample is drawn without replacement from the source file with a fixed seed (42) and kept in the drawn order, so row position is unrelated to the filter columns (``random-order'' layout; the file's original order is not used). We checked this on the full files: in source order, rows matching the query are strongly autocorrelated (lag-1 correlation of the match indicator 0.08 for taxi and 0.14 for flights), whereas in the samples it is within 0.002 of zero at every size, and the match rate in each twentieth of a sample stays within about 0.15 percentage points of the overall rate at 1 and 3 million rows. We also test a 1,000,000-row ``clustered'' layout, sorted by the filter columns, to evaluate index use and row-group skipping.

\subsection{Cold-cache protocol}
A unit of measurement writes the dataset to disk in one format, flushes it with \texttt{fsync}, evicts cached pages, and then times a read and a query. On macOS, eviction uses \texttt{sudo purge}, which empties the unified buffer cache. On the Linux host, which runs without root privileges in a small virtual machine, we call \texttt{posix\_fadvise(DONTNEED)} on every file belonging to the unit after \texttt{fsync}. Because a purge also evicts the code of the Python libraries, each unit first runs the same code path on a 64-row file, outside the timed region, so that library page-in is not charged to the large file.

Each measured unit therefore runs the following steps in order: (1) write a 64-row file in the format, outside the timed region; (2) write and \texttt{fsync} the full file, including any journal or sidecar files (for SQLite this includes creating the index and running \texttt{ANALYZE}, which are not part of retrieval time); (3) evict the cache, run the code path on the 64-row file, and time a full read into memory; (4) evict the cache again, prime again, and time the retrieval query, which opens the file or database afresh. Reads and retrieval are thus two independently cold operations on a file written moments earlier in the same unit.

We verify the protocol rather than assume it. At the start of every session, a probe file is read cold and warm. On the Mac the cold read ran at 4.6~GB/s and the warm read at 18.7~GB/s (cold 4.1 times slower); on the Linux host the figures were 5.3 and 16.5~GB/s (3.1 times). Every run also carries a flag raised if its read throughput exceeds a fixed ceiling of 8~GB/s, chosen as a conservative upper limit for the laptop's internal SSD; no run in any session was flagged. The fastest full reads were Pickle loads of files over 200~MB, which ran at 4.2 to 6.7~GB/s (median 5.6, 120 reads), and no read of any format exceeded 7~GB/s. Because probe throughput is a lower bound on what the drive can do, not a limit, we also checked the disk's own read counter. In one extra session on the main host we measured, for taxi data at 1 and 3 million rows and flight data at 3 million rows, the bytes the internal disk reported reading (the driver's separate read counter, not the combined read-and-write figure) during the timed full read and during the timed query, for Pickle, Parquet, Feather, CSV, ORC, SQLite and DuckDB, and for the Parquet projected and filtered scans, three repetitions each (144 measurements). After each measurement we repeated the same purge, priming read and pause, and recorded the counter over a window of the same length with no read, as an idle control. For the 99 operations that read a whole file (full reads and queries of Pickle, Parquet, Feather, CSV and ORC, and the DuckDB full read), the disk read between 0.91 and 1.34 of the file size (median 1.00; 75 of the 99 between 0.97 and 1.08, and the lowest 0.91), so the measured read volumes are consistent with predominantly device-backed reads under the checked conditions. Ratios above 1 occurred with the longer reads and are consistent with other disk traffic. SQLite queries read 1.00 to 1.07 of the file and SQLite full reads 1.19 to 1.49, which includes page overhead. The Parquet projected and filtered scans and the DuckDB query read 0.08 to 0.20 of the file (3.5 to 21.5~MB), which matches reading only the needed columns, and the counter showed device reads even for operations that finished in 8 to 20~ms. The idle control shows that other disk traffic is small but not zero: over windows shorter than 0.1~s the idle counter had a median of 0.4~MB (90th percentile 5.3~MB, maximum 15.8~MB), over 0.1 to 1~s a median of 1.9~MB (maximum 98~MB), and over the 1 to 21~s SQLite full reads a median of 25~MB (maximum 526~MB); writes during the timed windows did not exceed 4.7~MB. This is negligible against full-file reads of 27 to 1,087~MB but can matter for the partial scans, so for those we read the counter as evidence of device activity and not as an exact byte count. The check covers one session, the main host, and the sizes and datasets named above; we did not run it for the 50,000 and 250,000-row files, the synthetic data or the Linux host, where the cache is evicted per file, so for those we rely on the cache purge or eviction, the priming read, the session self-test and the throughput ceiling. A separate cold sequential-read test (single-threaded and with up to eight threads) reached at most 5.7~GB/s, below the Pickle reads, which confirms that simple sequential tests understate the drive's capability for this access pattern.

\subsection{Session design}
Within a session, all (configuration, repetition) units for a cell are shuffled, and cells themselves are executed in random order, which removes the confound of fixed format order and back-to-back repetition. The first unit of each configuration is a warm-up and is excluded. Random-order layouts use seven timed repetitions and the clustered layout five. The main Mac experiment has six sessions, three with the taxi and synthetic data and three with the synthetic and flight data, with the machine restarted before each and connected to AC power. Each session comprises 1,140 units with no errors. The secondary host has three sessions.

\subsection{Hosts}
The main host is a laptop with an Apple M4 Pro processor, 24~GB of memory, and macOS, running Python 3.12.9 with pandas 3.1, PyArrow 26 and DuckDB 2.0. The secondary host is a Linux virtual machine with 4~GB of memory and four cores, Python 3.10.12, pandas 2.3.3, PyArrow 25 and DuckDB 1.5.6. It cannot be restarted between sessions. The flight data is capped at 1,000,000 rows there, and each unit runs in its own process so that an out-of-memory kill is recorded as a result. To make older pandas behave like pandas 3, text columns are converted to Arrow strings, and the JSON Lines and SQLite writers operate in 100,000-row chunks on this host only. The two hosts therefore differ in operating system, memory, library versions, and cache-eviction method; differences between them cannot be attributed to memory alone.

\subsection{Statistics}
We report medians over all timed runs pooled across sessions. A 95\% interval for a median or a ratio of medians comes from a bootstrap that resamples runs within each session, keeping the sessions fixed, because three sessions are too few to resample. Per-session values are always available so between-session spread is visible. Pairwise performance differences are summarized using ratios of median retrieval times, accompanied by 95\% bootstrap confidence intervals. Ratios below one indicate that the first configuration is faster. Statistical significance is evaluated using within-session permutation tests, with Holm correction for multiple comparisons; Welch and Mann--Whitney tests serve as secondary checks. Bootstrap intervals are conditional on the observed sessions and do not account for uncertainty across unobserved sessions. The Holm correction~\cite{holm1979} is applied within families, each family consisting of all comparisons for one dataset and one metric; the comparisons are listed in Table~\ref{tab:contrasts}. We use 2,000 bootstrap and 10,000 permutation resamples. We do not report post hoc power~\cite{hoenig2001}.

The primary metric, \emph{retrieval time}, is the cold wall-clock time from a file at rest to the answer to the query. We also report write time and the time to read the whole file into memory.

\section{RESULTS}
\subsection{Overall ordering (RQ1)}
\begin{figure*}[t]
\centering
\includegraphics[width=\textwidth]{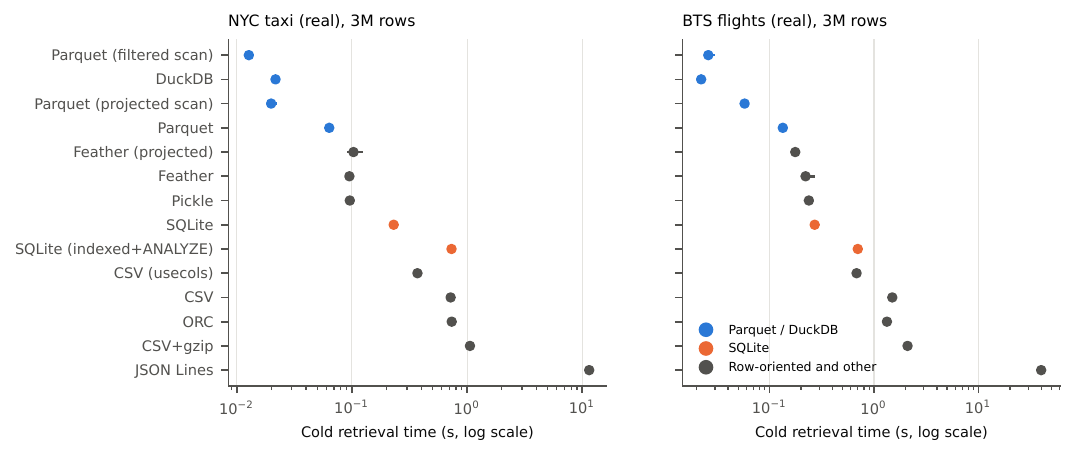}
\caption{Cold retrieval time per configuration at 3 million rows on the main host (median and 95\% bootstrap interval; intervals are narrower than the markers for most configurations). Log scale.}
\label{fig:retrieval}
\end{figure*}

\begin{figure*}[t]
\centering
\includegraphics[width=\textwidth]{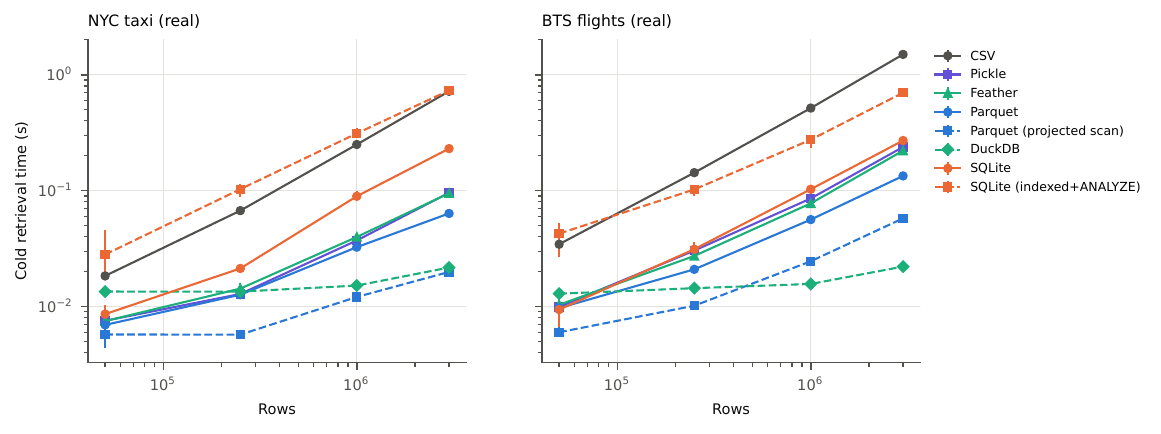}
\caption{Cold retrieval time as a function of row count for eight configurations, random-order layout, main host (median and 95\% interval). Log--log axes.}
\label{fig:scaling}
\end{figure*}

Fig.~\ref{fig:retrieval} shows cold retrieval time at 3 million rows, and Table~\ref{tab:retrieval} lists medians at 1 and 3 million rows. Fig.~\ref{fig:scaling} shows how the main configurations scale. On both datasets the fastest group is Parquet with a projected or filtered scan, DuckDB, and plain Parquet, followed by Feather, Pickle, and SQLite without an index, then indexed SQLite, CSV, and ORC, with compressed CSV and JSON Lines last. At 3 million rows on the taxi data, the cold query took 0.013~s with a Parquet filtered scan, against 0.72~s for CSV and 11.5~s for JSON Lines; on the flight data the figures were 0.026~s, 1.50~s and 39.8~s.

Table~\ref{tab:contrasts} summarizes pairwise contrasts as ranges of the ratio of medians over the four sizes, with the number of sizes that remain significant after Holm correction; the ranges describe how the effect changes with size and are not confidence intervals. Table~\ref{tab:contrast_ci} gives the ratio with its 95\% bootstrap interval and the corrected $p$ for each comparison at every size. Parquet needed 0.09 to 0.38 of the time of CSV (2.6 to 11 times faster), significant at all sizes on both datasets. JSON Lines was 10 to 27 times slower than CSV. Compressed CSV was 1.3 to 1.5 times slower than uncompressed CSV, so compression reduced file size to about a quarter but did not reduce cold retrieval time on this storage. Hence a single statement of ``orders of magnitude'' is not valid for every comparison; the ratio depends on the pair, the size, and the dataset, and we report the ranges.

\begin{table}[t]
\centering
\caption{Median cold retrieval time in seconds, random-order layout, main host}
\label{tab:retrieval}
\small
\setlength{\tabcolsep}{3.5pt}
\begin{tabular}{@{}lrrrr@{}}
\toprule
& \multicolumn{2}{c}{NYC taxi} & \multicolumn{2}{c}{BTS flights}\\
Configuration & 1M & 3M & 1M & 3M\\
\midrule
CSV & 0.250 & 0.719 & 0.515 & 1.50 \\
CSV (usecols) & 0.130 & 0.370 & 0.234 & 0.681 \\
CSV+gzip & 0.350 & 1.06 & 0.700 & 2.09 \\
JSON Lines & 3.80 & 11.5 & 9.83 & 39.8 \\
Pickle & 0.037 & 0.096 & 0.086 & 0.238 \\
Feather & 0.040 & 0.095 & 0.078 & 0.221 \\
Feather (projected) & 0.042 & 0.103 & 0.065 & 0.176 \\
Parquet & 0.033 & 0.064 & 0.056 & 0.134 \\
Parquet (projected scan) & 0.012 & 0.020 & 0.025 & 0.058 \\
Parquet (filtered scan) & 0.010 & 0.013 & 0.014 & 0.026 \\
Parquet (filtered scan, 100k row groups) & 0.009 & 0.012 & 0.012 & 0.023 \\
DuckDB & 0.015 & 0.022 & 0.016 & 0.022 \\
ORC & 0.288 & 0.735 & 0.489 & 1.33 \\
SQLite & 0.089 & 0.230 & 0.103 & 0.270 \\
SQLite (indexed+ANALYZE) & 0.311 & 0.732 & 0.275 & 0.699 \\

\bottomrule
\end{tabular}
\end{table}

\begin{table*}[t]
\centering
\caption{Retrieval-time contrasts, main host, random-order layout. For each pair (first configuration over second; a ratio below 1 means the first is faster): the range of the ratio of medians over 50k to 3M rows and the number of the four sizes significant after Holm correction. These ranges show how the observed ratio changes with size; they are not confidence intervals. The ratio and 95\% interval for each comparison at every size are in Table~\ref{tab:contrast_ci}.}
\label{tab:contrasts}
\small
\setlength{\tabcolsep}{4pt}
\begin{tabular}{@{}lcccc@{}}
\toprule
& \multicolumn{2}{c}{NYC taxi} & \multicolumn{2}{c}{BTS flights}\\
\cmidrule(lr){2-3}\cmidrule(lr){4-5}
Contrast & Ratio & Sig. & Ratio & Sig.\\
\midrule
Parquet $/$ CSV & 0.09--0.38 & 4/4 & 0.09--0.28 & 4/4 \\
JSON Lines $/$ CSV & 10.34--15.95 & 4/4 & 12.56--26.63 & 4/4 \\
CSV+gzip $/$ CSV & 1.31--1.47 & 4/4 & 1.27--1.40 & 4/4 \\
ORC $/$ Parquet & 3.23--11.56 & 4/4 & 3.60--9.93 & 4/4 \\
Pickle $/$ Parquet & 1.01--1.51 & 2/4 & 1.02--1.78 & 3/4 \\
Parquet $/$ Feather & 0.67--0.93 & 3/4 & 0.61--0.94 & 3/4 \\
CSV (usecols) $/$ CSV & 0.51--0.74 & 4/4 & 0.46--0.59 & 4/4 \\
Parquet projected $/$ full & 0.31--0.83 & 4/4 & 0.43--0.62 & 4/4 \\
Parquet filtered $/$ projected & 0.64--1.12 & 1/4 & 0.45--1.19 & 2/4 \\
DuckDB $/$ Parquet projected & 1.09--2.35 & 3/4 & 0.38--2.15 & 4/4 \\
SQLite indexed $/$ no index & 3.18--4.82 & 4/4 & 2.59--4.52 & 4/4 \\

\bottomrule
\end{tabular}
\end{table*}

\subsection{Text parsing versus binary decoding}
Earlier discussions of format performance often explain the advantage of binary formats by saying that they avoid deserializing the whole file. Our data do not support that explanation as stated. Pickle and plain Parquet both materialize the entire table before the pandas filter runs, and both were fast: at 3 million rows on the taxi data a full read took 0.084~s for Pickle and 0.058~s for Parquet, against 0.70~s for CSV and 11.7~s for JSON Lines. The slow tier is therefore formed by text formats, which pay for parsing, while the fast tier consists of binary formats that decode values with little parsing. Binary encoding alone is not sufficient, however: ORC, a binary columnar format, was 3.2 to 11.6 times slower than Parquet in the same pandas path on the taxi data and 3.6 to 9.9 times slower on the flight data, and at 3 million rows was as slow as CSV. We did not investigate the cause, which may lie in the reader implementation rather than the format, and the result applies to the PyArrow-to-pandas path we used.

Parquet and Feather were close. Parquet was 0.61 to 0.94 of Feather's retrieval time, significant at three of four sizes on each dataset, despite Feather files being larger (Tables~\ref{tab:retrieval_app} and~\ref{tab:contrast_ci}). Pickle was slower than Parquet by 1.0 to 1.8 times, significant at two to three of four sizes.

\subsection{Column projection and row-group skipping (RQ2)}
Reading only the needed columns is the largest single access-path gain within Parquet. A projected scan took 0.31 to 0.83 of the time of a full read on the taxi data and 0.43 to 0.62 on the flight data, significant at every size (Table~\ref{tab:contrasts}). The same idea applied to CSV (\texttt{usecols}) saved 26 to 54\% of the time; it must still tokenize every line.

\begin{figure*}[t]
\centering
\includegraphics[width=\textwidth]{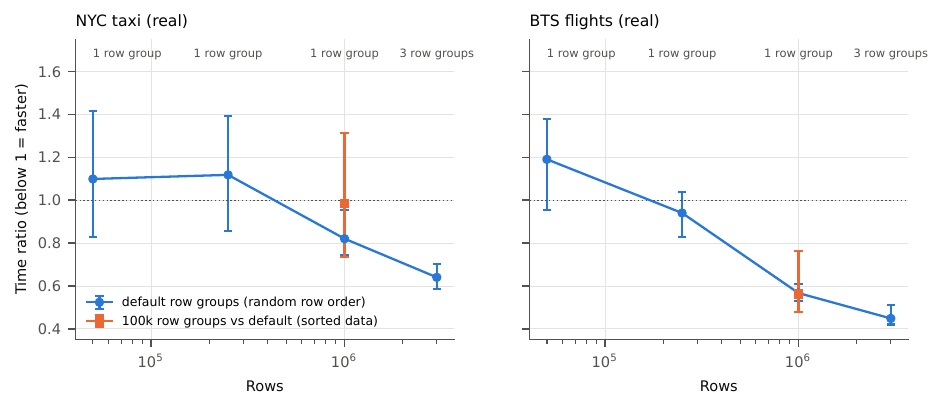}
\caption{Effect of row filtering inside the Parquet reader (blue, relative to a projected scan, random row order) and of 100k-row groups on sorted data (orange, relative to default row groups). Annotations give the number of row groups the default writer produced. Main host, medians and 95\% intervals.}
\label{fig:rowgroups}
\end{figure*}

Adding a row filter to the Parquet reader gave no benefit at 50,000 and 250,000 rows. Its time relative to a projected scan was 0.82 on taxi data and 0.57 on flight data at 1 million rows; only the flight difference was significant after Holm correction, so at that size we cannot claim a benefit on taxi data. At 3 million rows the ratios were 0.64 and 0.45, both significant. Row-group skipping cannot explain the gain at 1 million rows, because the default writer produced one row group at 50,000, 250,000 and 1,000,000 rows and three at 3,000,000 rows, with up to 1,048,576 rows each (confirmed with PyArrow 25 on the Linux host and PyArrow 26 on the Mac). With one row group there is nothing to skip, so the flight-data gain at that size, which is the only significant one, is not due to skipping and is consistent with filtering inside the reader before the table is converted to pandas; we did not isolate that mechanism. Skipping becomes possible only with several row groups. Writing 100,000-row groups (1, 3, 10 and 30 groups at the four sizes) was evaluated across sizes on randomly ordered data; on data sorted by the filter columns, tested at 1 million rows (10 groups), it reduced the filtered-scan time relative to the default layout to 0.56 and 0.59 of its value on flight and synthetic data (significant) but not on taxi data (0.98). On randomly ordered data, across all four sizes, the smaller groups gave no significant gain on either real dataset. Fig.~\ref{fig:rowgroups} shows both effects.

DuckDB behaved differently because it pays a fixed start-up cost. It was 1.4 to 2.7 times slower than a Parquet projected scan at 50,000 and 250,000 rows on flight and synthetic data, and 0.38 to 0.64 of the time (faster) at 1 and 3 million rows; on the taxi data it was slower or equal at all sizes (ratios 1.09 to 2.35) (Table~\ref{tab:contrasts}).

\subsection{SQLite secondary index (RQ3)}
On randomly ordered data, adding a composite index and running \texttt{ANALYZE} made the filtered query \emph{slower} than a full scan of the same SQLite table, by 3.2 to 4.8 times on taxi data and 2.6 to 4.5 times on flight data, significant at every size. On data sorted by the filter columns the index made the query about 8 to 10 times faster (ratio 0.10 to 0.12 at 1 million rows).

\begin{figure*}[t]
\centering
\includegraphics[width=\textwidth]{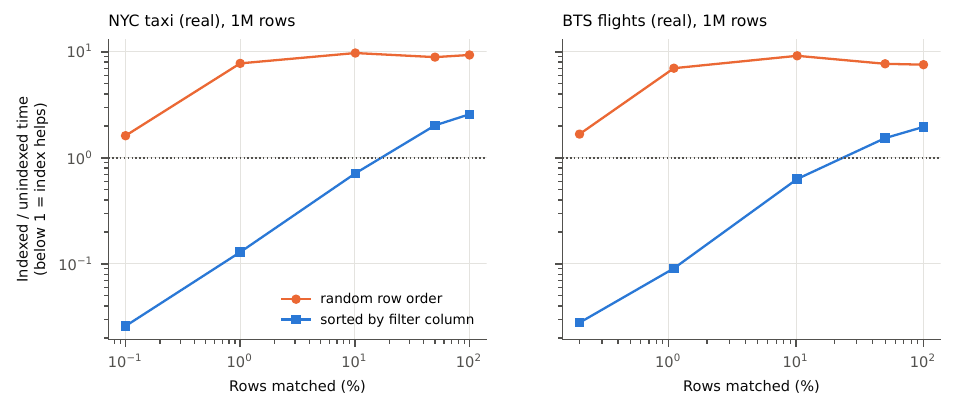}
\caption{Indexed SQLite query time relative to the unindexed time as a function of selectivity, for randomly ordered rows (orange) and rows sorted by the filter column (blue). Values below 1 mean the index helps. 1 million rows, main host.}
\label{fig:sqlite}
\end{figure*}

The selectivity sweep at 1 million rows explains this (Fig.~\ref{fig:sqlite}). It filters on a numeric column (taxi: total fare; flight: distance; three sessions each, not counting a first taxi session that used a different filter column), with actual selectivity from 0.1\% to 100\%. On randomly ordered data the indexed query was 1.6 to 1.7 times slower than the unindexed one even at 0.1 to 0.2\% selectivity, and 7 to 10 times slower from 1\% upward. On sorted data it was about 36 to 38 times faster at 0.1 to 0.2\% (ratio 0.03), 0.09 to 0.13 at 1\%, 0.63 to 0.71 at 10\%, and slower by 1.5 to 2.6 times at 50 to 100\%; the break-even therefore lies between the 10 and 50\% points we measured. The index is not a covering index, so every matching entry requires a lookup in the table to fetch the remaining columns. The results are consistent with scattered table reads on randomly ordered rows, which are expensive after a cache purge, and with a single contiguous range on sorted rows. We checked the planner's choice with \texttt{EXPLAIN QUERY PLAN} (SQLite 3.45.3, on the main host) for every combination of dataset, layout, selectivity and index condition, 72 plans in all, including the two-column query of the main experiment. Without an index the plan was always a full table scan. With the index, with or without \texttt{ANALYZE}, the planner chose a search through the index in all 48 cases, including at 50 and 100\% selectivity where a scan would read fewer pages, and it never reported a covering index. The plans were therefore identical with and without \texttt{ANALYZE}, which is consistent with the absence of a timing difference between those two conditions. The slowdown at high selectivity thus reflects the planner continuing to use a non-covering index when most rows match, not a change of plan. To test whether the statistics gathered by \texttt{ANALYZE} explain the result, we repeated the sweep with three SQLite conditions (no index, index with \texttt{ANALYZE}, index without \texttt{ANALYZE}) in three further sessions at 1 million rows (360 timed runs per session, no errors, no run flagged as impossibly fast). Omitting \texttt{ANALYZE} changed nothing we could detect: across the 20 combinations of dataset, layout and selectivity, the time of the index without \texttt{ANALYZE} relative to the index with it ranged from 0.89 to 1.32, and no difference was significant after Holm correction. The intervals were wide for sorted data at low selectivity (for example 0.57 to 2.05 for flights at 0.2\%), so we report absence of a detected difference, not proof of equivalence. The index effects against the unindexed table reproduced the first sweep: 1.4 to 1.5 times slower on random order at 0.1 to 0.2\% selectivity and 6.6 to 9.1 times slower from 1\% upward; on sorted data 0.03 to 0.04 of the unindexed time at 0.1 to 0.2\%, 0.07 to 0.13 at 1\%, 0.58 to 0.69 at 10\%, and 1.7 to 2.4 at 50 to 100\%.

\subsection{Stability across sessions and datasets (RQ4)}
Between-session variability was modest. The median coefficient of variation of the per-session median retrieval time was 6\% on the main host (5.2\% flights, 5.8\% taxi, 7.0\% synthetic) and 3\% on the Linux host; the largest variability, about 60\%, occurred for indexed SQLite. After Holm correction only one of 225 retrieval cells on the main host and none of 124 on the Linux host showed a significant session effect (Kruskal--Wallis); with three sessions this test has little power, so we read the absence of significance as consistent with, not proof of, stable sessions, and report per-session ranges instead. The synthetic dataset, which was run in two independent batches, gave consistent medians (Spearman $\rho = 0.99$ between batches), supporting repeatability of the protocol.

The two real datasets agree on the tiers but not always on effect sizes; for example the Parquet filtered-scan benefit at 3 million rows was 0.64 on the taxi data and 0.45 on the flight data, and DuckDB compared with a Parquet projected scan was slower on the taxi data at 3 million rows and faster on flight data. We therefore describe the tiers as qualitatively similar with different magnitudes.

\subsection{Write time and storage size}
The 3-million-row medians in Table~\ref{tab:retrieval_app} show that the formats that read fastest were also cheap to write and compact, with exceptions. Parquet files were 81~MB (taxi) and 113~MB (flight), about a quarter and a fifth of the 329 and 586~MB of the CSV files, and writing them took 0.81 and 1.5~s against 9.0 and 14.9~s for CSV, 11 and 10 times faster. Feather and Pickle were fastest to write (0.09 and 0.30~s on taxi data) but produced files two to ten times larger than Parquet (153 and 475~MB on taxi data, 377~MB and 1.09~GB on flight data). Gzip-compressed CSV was smallest on taxi data (74~MB) but took 37 and 114~s to write, 4.2 and 7.6 times the plain CSV time, and its cold retrieval was slower than plain CSV. JSON Lines wrote slightly faster than CSV (6.7 and 12.7~s) but produced files 4.4 and 4.5 times larger (1.4 and 2.6~GB). ORC was about the size of CSV and wrote quickly. SQLite wrote at roughly CSV speed (7.4 and 10.4~s); adding the index and running \texttt{ANALYZE} raised the write time by 15 and 13\% and the file size by 10 and 11\%, which, together with the retrieval results, means the index was a cost on randomly ordered data. Writing 100,000-row groups increased the Parquet file by 13\% on taxi data and 1.3\% on flight data and the write time by 20 and 6\%. Write times are single warm writes of an in-memory table, including the flush to disk, repeated across runs and sessions; they were not measured under a cold protocol, and they depend on the pandas and PyArrow writers rather than on the format alone. Table~\ref{tab:retrieval_app} gives the 95\% interval for each write-time median and, separately, for each cold-retrieval median.

\subsection{Small-memory host}
The 4~GB host completed all sizes for the flight data up to its one-million-row cap and the taxi data up to 3 million, but with out-of-memory kills for some configurations that completed on the Mac (Table~\ref{tab:vm}). The kills were identical in all three sessions: JSON Lines at 1 million rows and above on both datasets, and SQLite with and without an index at 1 million rows on flight data and 3 million rows on taxi data.

\begin{table}[t]
\centering
\caption{Configurations that ran out of memory on the 4~GB host in all three sessions (all completed on the 24~GB host)}
\label{tab:vm}
\small
\begin{tabular}{@{}lll@{}}
\toprule
Configuration & NYC taxi & BTS flights\\
\midrule
JSON Lines & 1M and 3M rows & 1M rows\\
SQLite, SQLite (indexed) & 3M rows & 1M rows\\
\bottomrule
\end{tabular}
\end{table}

\begin{figure}[t]
\centering
\includegraphics[width=\columnwidth]{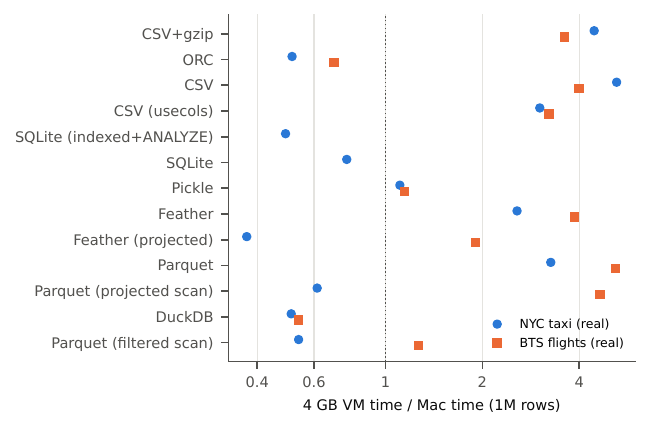}
\caption{Ratio of cold retrieval time on the 4~GB Linux host to the Mac at 1 million rows, random-order layout (above 1 means slower on the Linux host). Configurations that ran out of memory are omitted.}
\label{fig:vm}
\end{figure}

For the configurations that completed, the ratio of VM time to Mac time ranged from about 0.4 to 5 at 1 million rows depending on the configuration (Fig.~\ref{fig:vm}), so the Linux host was not uniformly slower: CSV, plain Parquet, and Feather were 2.5 to 5 times slower, whereas DuckDB, ORC and indexed SQLite were faster. Rankings agreed broadly (Spearman 0.80 to 0.92 per cell across 12 to 15 configurations), but some pairwise orderings reversed. Pickle was 1.0 to 1.8 times \emph{slower} than Parquet on the Mac and 0.33 to 0.42 of Parquet's time (that is, about 2.4 to 3 times \emph{faster}) on the Linux host, a significant reversal on both datasets. ORC was 1.2 to 1.4 times slower than Parquet on the Linux host, against 3 to 12 times on the Mac. A filtered Parquet scan reduced time relative to a projected scan at every size on the Linux host (0.16 to 0.77), against inconsistent gains on the Mac. The SQLite index remained slower on randomly ordered data (1.7 to 2.3 times). Because the two hosts differ in operating system, memory, library versions and eviction method, we do not attribute any of these differences to memory alone.

\section{DISCUSSION}
\textbf{Practical guidance.} For analytical filtering of a file read once from storage, a Parquet file with a projected or filtered scan, or DuckDB over the same data, was the fastest path in our tests at one million rows and above, and a Parquet projected scan was fastest at smaller sizes (Fig.~\ref{fig:scaling}). At 3 million rows, JSON Lines was 10 to 27 times slower than CSV and up to roughly three orders of magnitude slower than the fastest path. If the data must stay in a relational store, our results show that a non-covering secondary index should be evaluated against the physical order of the table and the selectivity of the query rather than assumed to help: on randomly ordered rows it made cold queries slower in every case we tested, and on sorted rows it helped at 10\% selectivity and below but not at 50\% and above. We tested one index design and one SQLite version per host, and other index designs, such as covering indexes, may behave differently. A row filter inside the Parquet reader is worth trying from about one million rows; its gain at that size was significant on flight data but not on taxi data. To benefit from row-group skipping, files must hold several row groups, which the default writer produced only at 3 million rows, and the data must be organized so that filter columns correlate with row order.

\textbf{On cold versus cached measurements.} Our protocol measures first access from storage. Workloads that repeatedly access the same file see cached performance, which our setup excludes by design. The probe results show that cold reads cost three to four times more than warm reads on these machines. Studies that do not control the cache therefore describe a different regime, and the choice between regimes should match the intended workload.

\textbf{Reproducing orderings.} The reversal of Pickle and Parquet on the smaller host illustrates that format rankings depend on more than the format. A practitioner choosing between two formats with ratios near one should test on the target environment.

\section{LIMITATIONS}
This study has several limitations. (1) Each dataset has one query for the main experiment; the sweep varies selectivity on a different filter column but does not cover joins, multi-column aggregates, or concurrent workloads. (2) The consumer is pandas through PyArrow; other engines would change the results for ORC and others. (3) Three sessions per dataset are too few to estimate the session-level variance precisely. (4) The two hosts differ in several respects, and the Linux host cannot be restarted between sessions and uses per-file eviction instead of a full cache purge, so cross-host differences are descriptive. (5) Both datasets are single-table data with a moderate number of columns; results may differ for wide or nested schemas. (6) The bootstrap interval for a ratio of medians and the permutation p-value can disagree in borderline cells; we report both. (7) We measured wall-clock time only, not memory or energy. (8) Results depend on library versions; we record them in the released environment files.

\section{CONCLUSION}
Under an explicitly cold cache, a multi-session design, and two real datasets, columnar files read through projected or filtered scans, and DuckDB, answered a filtered aggregation faster than the other configurations from one million rows upward; text formats were far slower, and the evaluated non-covering secondary index slowed SQLite queries on the tested random-order layouts. A reader-side row filter was significantly faster than a projected scan on flight data from one million rows and on both datasets at three million, row-group skipping needs several row groups (the default writer produced them only at 3 million rows), and pairwise orderings such as Pickle against Parquet changed on a smaller host. We release code, data derivation and raw timings so that others can repeat the measurements with their own hardware.

\section*{DATA AND CODE AVAILABILITY}
The benchmark scripts, analysis scripts, raw timing files, and environment records are available on Zenodo at \url{https://doi.org/10.5281/zenodo.23927097}. The two real datasets are public: NYC TLC trip records~\cite{tlc} and BTS on-time data~\cite{bts}; the derivation of each sample is scripted and identified by file hashes in Appendix~D.

\section*{ACKNOWLEDGMENT}
The first author used the AI assistant Claude (Anthropic) to help generate the figures, to help draft the manuscript, and to help write the code for the analysis.

\appendices
\clearpage\onecolumn
\section{Detailed results at 3 million rows}
\begin{table}[H]
\centering
\caption{Output size, median write time with 95\% interval, median full-read time, and median cold retrieval time with 95\% interval at 3 million rows, random-order layout, main host (seconds unless noted)}
\label{tab:retrieval_app}
\footnotesize
\renewcommand{\arraystretch}{0.82}
\begin{tabular}{@{}llrrcrrc@{}}
\toprule
Dataset & Configuration & Size (MB) & Write & Write 95\% interval & Read & Retrieve & Retrieve 95\% interval\\
\midrule
NYC & CSV & 329 & 8.96 & [8.92, 8.99] & 0.698 & 0.719 & [0.710, 0.730] \\
NYC & CSV (usecols) & 329 & 8.91 & [8.88, 8.94] & 0.688 & 0.370 & [0.369, 0.375] \\
NYC & CSV+gzip & 74 & 37.4 & [37.4, 37.5] & 1.02 & 1.06 & [1.01, 1.07] \\
NYC & JSON Lines & 1437 & 6.69 & [6.68, 6.71] & 11.7 & 11.5 & [11.3, 11.5] \\
NYC & Pickle & 475 & 0.300 & [0.218, 0.327] & 0.084 & 0.096 & [0.095, 0.098] \\
NYC & Feather & 153 & 0.094 & [0.092, 0.095] & 0.080 & 0.095 & [0.089, 0.096] \\
NYC & Feather (projected) & 153 & 0.093 & [0.093, 0.095] & 0.080 & 0.103 & [0.093, 0.121] \\
NYC & Parquet & 81 & 0.805 & [0.803, 0.812] & 0.058 & 0.064 & [0.063, 0.066] \\
NYC & Parquet (projected scan) & 81 & 0.805 & [0.801, 0.809] & 0.057 & 0.020 & [0.019, 0.022] \\
NYC & Parquet (filtered scan) & 81 & 0.811 & [0.804, 0.812] & 0.055 & 0.013 & [0.012, 0.014] \\
NYC & Parquet (filtered scan, 100k row groups) & 91 & 0.969 & [0.965, 0.973] & 0.057 & 0.012 & [0.012, 0.014] \\
NYC & DuckDB & 98 & 0.864 & [0.860, 0.867] & 0.499 & 0.022 & [0.020, 0.023] \\
NYC & ORC & 336 & 0.710 & [0.704, 0.715] & 0.759 & 0.735 & [0.703, 0.762] \\
NYC & SQLite & 372 & 7.39 & [7.35, 7.43] & 7.31 & 0.230 & [0.215, 0.239] \\
NYC & SQLite (indexed+ANALYZE) & 408 & 8.51 & [8.49, 8.55] & 7.33 & 0.732 & [0.717, 0.785] \\
\hline
BTS & CSV & 586 & 14.9 & [14.9, 15.0] & 1.44 & 1.50 & [1.49, 1.51] \\
BTS & CSV (usecols) & 586 & 14.9 & [14.9, 14.9] & 1.44 & 0.681 & [0.671, 0.688] \\
BTS & CSV+gzip & 151 & 113.7 & [113.5, 114.1] & 2.04 & 2.09 & [2.01, 2.12] \\
BTS & JSON Lines & 2639 & 12.7 & [12.5, 14.0] & 42.1 & 39.8 & [39.4, 41.9] \\
BTS & Pickle & 1087 & 0.596 & [0.452, 0.695] & 0.194 & 0.238 & [0.233, 0.241] \\
BTS & Feather & 377 & 0.279 & [0.213, 0.432] & 0.226 & 0.221 & [0.213, 0.265] \\
BTS & Feather (projected) & 377 & 0.192 & [0.187, 0.198] & 0.174 & 0.176 & [0.164, 0.184] \\
BTS & Parquet & 113 & 1.49 & [1.48, 1.49] & 0.114 & 0.134 & [0.128, 0.135] \\
BTS & Parquet (projected scan) & 113 & 1.48 & [1.47, 1.49] & 0.105 & 0.058 & [0.057, 0.060] \\
BTS & Parquet (filtered scan) & 113 & 1.48 & [1.48, 1.49] & 0.109 & 0.026 & [0.024, 0.029] \\
BTS & Parquet (filtered scan, 100k row groups) & 115 & 1.57 & [1.57, 1.62] & 0.111 & 0.023 & [0.020, 0.028] \\
BTS & DuckDB & 206 & 1.57 & [1.56, 1.59] & 1.26 & 0.022 & [0.021, 0.023] \\
BTS & ORC & 530 & 1.40 & [1.39, 1.41] & 1.39 & 1.33 & [1.29, 1.35] \\
BTS & SQLite & 407 & 10.4 & [10.4, 10.5] & 21.4 & 0.270 & [0.262, 0.281] \\
BTS & SQLite (indexed+ANALYZE) & 453 & 11.8 & [11.7, 11.9] & 21.4 & 0.699 & [0.682, 0.761] \\

\bottomrule
\end{tabular}
\end{table}

\section{Small-memory host results}
\begin{table}[H]
\centering
\caption{Cold retrieval time in seconds on the 4~GB Linux host (three sessions): median over all runs, 95\% bootstrap interval, range of the three session medians in parentheses, and ratio of the Linux median to the Mac median (above 1 means slower on Linux). ``oom'' marks configurations killed for lack of memory in every session. Random-order layout.}
\label{tab:vm_full}
\footnotesize
\renewcommand{\arraystretch}{0.82}
\setlength{\tabcolsep}{3pt}
\resizebox{\textwidth}{!}{\begin{tabular}{@{}l cc cc cc@{}}
\toprule
& \multicolumn{2}{c}{NYC taxi, 1M rows} & \multicolumn{2}{c}{BTS flights, 1M rows} & \multicolumn{2}{c}{NYC taxi, 3M rows}\\
\cmidrule(lr){2-3}\cmidrule(lr){4-5}\cmidrule(lr){6-7}
Configuration & Median [95\% CI] (sessions) & Ratio & Median [95\% CI] (sessions) & Ratio & Median [95\% CI] (sessions) & Ratio\\
\midrule
CSV & 1.31 [1.29, 1.32] (1.29--1.32) & 5.22 & 2.06 [2.04, 2.07] (2.04--2.13) & 3.99 & 4.29 [4.28, 4.35] (4.26--4.36) & 5.96 \\
CSV (usecols) & 0.392 [0.389, 0.397] (0.385--0.395) & 3.02 & 0.755 [0.750, 0.758] (0.750--0.770) & 3.22 & 1.12 [1.11, 1.12] (1.10--1.12) & 3.01 \\
CSV+gzip & 1.56 [1.53, 1.57] (1.51--1.57) & 4.45 & 2.52 [2.51, 2.52] (2.49--2.61) & 3.59 & 4.98 [4.95, 5.10] (4.95--5.20) & 4.70 \\
JSON Lines & oom & -- & oom & -- & oom & -- \\
Pickle & 0.041 [0.040, 0.044] (0.040--0.044) & 1.11 & 0.098 [0.098, 0.100] (0.093--0.102) & 1.15 & 0.115 [0.113, 0.116] (0.113--0.116) & 1.20 \\
Feather & 0.102 [0.102, 0.103] (0.102--0.104) & 2.56 & 0.300 [0.299, 0.311] (0.299--0.315) & 3.87 & 0.307 [0.307, 0.310] (0.300--0.336) & 3.24 \\
Feather (projected) & 0.016 [0.015, 0.016] (0.015--0.016) & 0.37 & 0.124 [0.120, 0.125] (0.120--0.125) & 1.91 & 0.045 [0.045, 0.046] (0.045--0.046) & 0.44 \\
Parquet & 0.106 [0.106, 0.109] (0.106--0.111) & 3.26 & 0.292 [0.288, 0.293] (0.288--0.299) & 5.19 & 0.308 [0.305, 0.309] (0.305--0.309) & 4.85 \\
Parquet (projected scan) & 0.007 [0.007, 0.008] (0.007--0.008) & 0.61 & 0.114 [0.112, 0.115] (0.111--0.115) & 4.63 & 0.019 [0.018, 0.019] (0.018--0.020) & 0.94 \\
Parquet (filtered scan) & 0.005 [0.005, 0.006] (0.005--0.006) & 0.54 & 0.018 [0.017, 0.018] (0.017--0.018) & 1.27 & 0.013 [0.012, 0.013] (0.012--0.013) & 1.00 \\
Parquet (filtered scan, 100k row groups) & 0.006 [0.005, 0.006] (0.005--0.006) & 0.61 & 0.013 [0.012, 0.014] (0.012--0.014) & 1.10 & 0.014 [0.013, 0.015] (0.013--0.015) & 1.10 \\
DuckDB & 0.008 [0.008, 0.008] (0.008--0.008) & 0.51 & 0.008 [0.008, 0.009] (0.008--0.009) & 0.54 & 0.013 [0.013, 0.013] (0.013--0.013) & 0.60 \\
ORC & 0.148 [0.147, 0.150] (0.147--0.150) & 0.51 & 0.339 [0.337, 0.340] (0.330--0.346) & 0.69 & 0.436 [0.435, 0.449] (0.435--0.441) & 0.59 \\
SQLite & 0.068 [0.067, 0.069] (0.067--0.069) & 0.76 & oom & -- & oom & -- \\
SQLite (indexed+ANALYZE) & 0.152 [0.150, 0.156] (0.150--0.156) & 0.49 & oom & -- & oom & -- \\

\bottomrule
\end{tabular}}
\end{table}

\section{Comparison intervals at every size}
\begin{table}[H]
\centering
\caption{Retrieval-time ratio of medians with 95\% bootstrap interval (resampling runs within sessions) and Holm-corrected permutation $p$ (values below 0.01 are displayed as $<$0.01). Significance is judged from the Holm-adjusted $p$, not from whether the unadjusted interval excludes 1, so an interval that excludes 1 can accompany a non-significant adjusted $p$ (for example taxi, 1M rows, filtered versus projected scan) for the contrasts of Table~\ref{tab:contrasts}, at each size. First configuration over second; main host, random-order layout.}
\label{tab:contrast_ci}
\footnotesize
\renewcommand{\arraystretch}{0.7}
\setlength{\tabcolsep}{3pt}
\begin{tabular}{@{}l cc cc cc cc@{}}
\toprule
Contrast & \multicolumn{2}{c}{50k rows} & \multicolumn{2}{c}{250k rows} & \multicolumn{2}{c}{1M rows} & \multicolumn{2}{c}{3M rows}\\
\cmidrule(lr){2-3}\cmidrule(lr){4-5}\cmidrule(lr){6-7}\cmidrule(lr){8-9}
& Ratio [95\% CI] & $p$ & Ratio [95\% CI] & $p$ & Ratio [95\% CI] & $p$ & Ratio [95\% CI] & $p$\\
\midrule
\multicolumn{9}{@{}l}{\emph{NYC taxi}}\\
Parquet $/$ CSV & 0.378 [0.309, 0.417] & $<$0.01 & 0.188 [0.182, 0.206] & $<$0.01 & 0.130 [0.126, 0.132] & $<$0.01 & 0.088 [0.087, 0.092] & $<$0.01 \\
JSON Lines $/$ CSV & 10.3 [9.13, 10.9] & $<$0.01 & 14.3 [14.2, 14.6] & $<$0.01 & 15.2 [15.0, 15.3] & $<$0.01 & 16.0 [15.7, 16.2] & $<$0.01 \\
CSV+gzip $/$ CSV & 1.31 [1.12, 1.35] & $<$0.01 & 1.32 [1.29, 1.40] & $<$0.01 & 1.40 [1.33, 1.44] & $<$0.01 & 1.47 [1.40, 1.50] & $<$0.01 \\
ORC $/$ Parquet & 3.23 [2.75, 4.22] & $<$0.01 & 7.06 [6.02, 8.33] & $<$0.01 & 8.86 [8.49, 9.38] & $<$0.01 & 11.6 [10.9, 12.1] & $<$0.01 \\
Pickle $/$ Parquet & 1.09 [0.847, 1.34] & 1.00 & 1.01 [0.937, 1.08] & 1.00 & 1.14 [1.09, 1.20] & $<$0.01 & 1.51 [1.46, 1.56] & $<$0.01 \\
Parquet $/$ Feather & 0.927 [0.693, 1.03] & 0.84 & 0.886 [0.801, 0.963] & 0.02 & 0.818 [0.770, 0.847] & $<$0.01 & 0.669 [0.655, 0.725] & $<$0.01 \\
CSV (usecols) $/$ CSV & 0.739 [0.651, 0.861] & $<$0.01 & 0.539 [0.530, 0.554] & $<$0.01 & 0.520 [0.510, 0.526] & $<$0.01 & 0.515 [0.506, 0.524] & $<$0.01 \\
Parquet projected $/$ full & 0.827 [0.631, 0.975] & 0.02 & 0.453 [0.414, 0.497] & $<$0.01 & 0.372 [0.351, 0.407] & $<$0.01 & 0.313 [0.300, 0.341] & $<$0.01 \\
Parquet filtered $/$ projected & 1.10 [0.829, 1.42] & 1.00 & 1.12 [0.857, 1.39] & 1.00 & 0.821 [0.742, 0.955] & 0.17 & 0.641 [0.587, 0.701] & $<$0.01 \\
DuckDB $/$ Parquet projected & 2.35 [2.15, 3.14] & $<$0.01 & 2.35 [2.09, 2.60] & $<$0.01 & 1.25 [1.14, 1.40] & $<$0.01 & 1.09 [0.994, 1.16] & 1.00 \\
SQLite indexed $/$ no index & 3.27 [2.16, 5.56] & $<$0.01 & 4.82 [3.97, 5.52] & $<$0.01 & 3.48 [3.04, 3.95] & $<$0.01 & 3.18 [3.01, 3.47] & $<$0.01 \\

\midrule
\multicolumn{9}{@{}l}{\emph{BTS flights}}\\
Parquet $/$ CSV & 0.282 [0.256, 0.316] & $<$0.01 & 0.147 [0.138, 0.153] & $<$0.01 & 0.109 [0.108, 0.112] & $<$0.01 & 0.089 [0.086, 0.091] & $<$0.01 \\
JSON Lines $/$ CSV & 12.6 [11.6, 13.1] & $<$0.01 & 17.5 [17.3, 17.7] & $<$0.01 & 19.1 [19.0, 19.2] & $<$0.01 & 26.6 [26.2, 28.0] & $<$0.01 \\
CSV+gzip $/$ CSV & 1.28 [1.17, 1.36] & $<$0.01 & 1.27 [1.23, 1.31] & $<$0.01 & 1.36 [1.34, 1.42] & $<$0.01 & 1.40 [1.33, 1.42] & $<$0.01 \\
ORC $/$ Parquet & 3.60 [3.03, 4.79] & $<$0.01 & 6.73 [6.36, 7.42] & $<$0.01 & 8.69 [8.25, 8.93] & $<$0.01 & 9.93 [9.60, 10.4] & $<$0.01 \\
Pickle $/$ Parquet & 1.02 [0.883, 1.15] & 1.00 & 1.45 [1.36, 1.60] & $<$0.01 & 1.52 [1.48, 1.57] & $<$0.01 & 1.78 [1.72, 1.85] & $<$0.01 \\
Parquet $/$ Feather & 0.937 [0.819, 1.10] & 1.00 & 0.766 [0.710, 0.815] & $<$0.01 & 0.725 [0.699, 0.745] & $<$0.01 & 0.606 [0.496, 0.630] & $<$0.01 \\
CSV (usecols) $/$ CSV & 0.586 [0.531, 0.661] & $<$0.01 & 0.472 [0.453, 0.489] & $<$0.01 & 0.455 [0.447, 0.462] & $<$0.01 & 0.455 [0.446, 0.460] & $<$0.01 \\
Parquet projected $/$ full & 0.618 [0.539, 0.760] & $<$0.01 & 0.486 [0.454, 0.532] & $<$0.01 & 0.437 [0.413, 0.449] & $<$0.01 & 0.431 [0.424, 0.459] & $<$0.01 \\
Parquet filtered $/$ projected & 1.19 [0.956, 1.38] & 1.00 & 0.941 [0.831, 1.04] & 1.00 & 0.567 [0.528, 0.611] & $<$0.01 & 0.449 [0.421, 0.511] & $<$0.01 \\
DuckDB $/$ Parquet projected & 2.15 [1.75, 2.37] & $<$0.01 & 1.42 [1.27, 1.53] & $<$0.01 & 0.637 [0.605, 0.677] & $<$0.01 & 0.384 [0.361, 0.405] & $<$0.01 \\
SQLite indexed $/$ no index & 4.52 [2.45, 6.54] & $<$0.01 & 3.28 [2.62, 3.73] & $<$0.01 & 2.67 [2.19, 3.16] & $<$0.01 & 2.59 [2.47, 2.82] & $<$0.01 \\

\bottomrule
\end{tabular}
\end{table}

\section{Reproducibility details}
\footnotesize
\textbf{Inputs.} The analysed files are the NYC TLC yellow-taxi trip file for June 2026 (3,837,248 rows, 65{,}465{,}637 bytes) and the BTS on-time file for January to July 2026 (4,120,343 rows, 43 columns, 139{,}911{,}910 bytes), both stored as Parquet. SHA-256 of the taxi file and of the flight file:

{\scriptsize\ttfamily\noindent
f340c62a18885c56176433c9db28e97d\\
0a8a14e6e49869276ad8afb20ebf088b\\[3pt]
06f6173918e2bea7a666e92074f5fdb0\\
b25fbff2d36381ea3fc809a529e4b098\par}

A sample of $n$ rows is the first $n$ entries of \texttt{numpy.random.RandomState(42).choice(N, size=n, replace=False)} applied to the file's rows in file order, with $N$ the file's row count. The leading 16 hexadecimal digits of the SHA-256 of the sampled row indices (as little-endian 64-bit integers) at $n=$ 50k, 250k, 1M and 3M are \texttt{eacb2ec71bdf96ff}, \texttt{708f10d251473676}, \texttt{26f05b01040cac59}, \texttt{bb15150b52473bbc} (taxi) and \texttt{14d27db08ed73a14}, \texttt{06e625a14eac06a0}, \texttt{831898a206ec1caf}, \texttt{2444e8148d4ae4d0} (flights). The clustered layout is the 1M sample sorted stably by the two filter columns. The synthetic control is generated from seed 42 by the same scripts.

\textbf{Queries.} The taxi query is \texttt{SELECT VendorID, SUM(fare\_amount) FROM t WHERE payment\_type = 2 AND PULocationID = 132 GROUP BY VendorID} (1.05\% of rows). The flight query is \texttt{SELECT DAY\_OF\_WEEK, SUM(ARR\_DELAY) FROM t WHERE OP\_UNIQUE\_CARRIER = 'UA' AND DEST = 'SFO' GROUP BY DAY\_OF\_WEEK} (0.88\%). The indexed SQLite configurations run \texttt{CREATE INDEX idx ON t(payment\_type, PULocationID)} (taxi) or on the two flight filter columns, followed by \texttt{ANALYZE} except in the ablation condition. The selectivity sweep uses \texttt{WHERE c <= x} with $c$ the total fare (taxi) or distance (flights) and $x$ the quantile that gives the target selectivity. The sweep indexes that filter column alone, \texttt{t(total\_amount)} (taxi) or \texttt{t(DISTANCE)} (flights), not the composite index of the main query. The same shape of query is implemented with pandas boolean masks, PyArrow filters, and DuckDB SQL, and every run records the sum and number of groups of its answer so that the configurations can be checked against each other.

\textbf{Software and environment.} Each session writes a JSON record with the platform, Python, pandas, NumPy, PyArrow and DuckDB versions, CPU, memory, power and thermal state, the cold-cache self-test, and the configuration (sizes, repetitions, warm-up, seed); versions are in Section~III-E and SQLite was 3.45.3 on the main host.

\textbf{Cache checks and results files.} Each session records the self-test (a 512~MB file, cold at least 1.3 times slower than warm and below 8~GB/s); every run records the cache method, a verification flag, the priming read, and a flag for throughput above 8~GB/s, together with session, dataset, layout, size, configuration, repetition, times, file size and answer. The device-counter check with its idle control, the ceiling test and the 72 query plans are released as separate files.

\twocolumn

\begin{IEEEbiographynophoto}{ASRA FATHIMA}
is pursuing the Ph.D.\ degree in information technology with a focus on data science at the University of the Cumberlands, Williamsburg, KY, USA. She received the M.S.\ degree in computer science from Oklahoma City University, Oklahoma City, OK, USA, and the B.Com.\ degree with a computer specialization from Pragathi Women's Degree College, India. She has worked as a software developer at a Fortune 500 company since 2022. Her research interests include data engineering, storage formats, and performance evaluation of data systems, and her current study measures cold-cache query performance of tabular data formats.
\end{IEEEbiographynophoto}

\begin{IEEEbiographynophoto}{KRISTOFFER ROBERTS}
is a professor at the University of the Cumberlands, Williamsburg, KY, USA. He holds a Ph.D. in information technology from Capella University and has worked in information technology for more than 20 years and has taught in person and online for nearly a decade. He holds several industry certifications, including Cisco Certified Network Professional (CCNP), Certified Information Systems Security Professional (CISSP), and Project Management Professional (PMP).
\end{IEEEbiographynophoto}

\begin{thebibliography}{00}
\bibitem{zeng2023} X. Zeng, Y. Hui, J. Shen, A. Pavlo, W. McKinney, and H. Zhang, ``An empirical evaluation of columnar storage formats,'' \textit{Proc. VLDB Endow.}, vol. 17, no. 2, pp. 148--161, 2023, doi: 10.14778/3626292.3626298.
\bibitem{plase2017} D. Plase, L. Niedrite, and R. Taranovs, ``A comparison of HDFS compact data formats: Avro versus Parquet,'' \textit{Mokslas -- Lietuvos ateitis / Sci. -- Future of Lithuania}, vol. 9, no. 3, pp. 267--276, 2017, doi: 10.3846/mla.2017.1033.
\bibitem{abadi2008} D. J. Abadi, S. R. Madden, and N. Hachem, ``Column-stores vs. row-stores: How different are they really?'' in \textit{Proc. ACM SIGMOD Int. Conf. Manage. Data}, 2008, pp. 967--980.
\bibitem{raasveldt2019} M. Raasveldt and H. M\"uhleisen, ``DuckDB: An embeddable analytical database,'' in \textit{Proc. ACM SIGMOD Int. Conf. Manage. Data}, 2019, pp. 1981--1984.
\bibitem{mckinney2010} W. McKinney, ``Data structures for statistical computing in Python,'' in \textit{Proc. 9th Python Sci. Conf.}, 2010, pp. 56--61.
\bibitem{georges2007} A. Georges, D. Buytaert, and L. Eeckhout, ``Statistically rigorous Java performance evaluation,'' in \textit{Proc. ACM OOPSLA}, 2007, pp. 57--76.
\bibitem{mytkowicz2009} T. Mytkowicz, A. Diwan, M. Hauswirth, and P. F. Sweeney, ``Producing wrong data without doing anything obviously wrong!'' in \textit{Proc. ACM ASPLOS}, 2009, pp. 265--276.
\bibitem{kalibera2013} T. Kalibera and R. Jones, ``Rigorous benchmarking in reasonable time,'' in \textit{Proc. ACM ISMM}, 2013, pp. 63--74.
\bibitem{hoefler2015} T. Hoefler and R. Belli, ``Scientific benchmarking of parallel computing systems: Twelve ways to tell the masses when reporting performance results,'' in \textit{Proc. Int. Conf. High Perform. Comput., Netw., Storage Anal. (SC)}, 2015, doi: 10.1145/2807591.2807644.
\bibitem{maricq2018} A. Maricq, D. Duplyakin, I. Jimenez, C. Maltzahn, R. Stutsman, and R. Ricci, ``Taming performance variability,'' in \textit{Proc. 13th USENIX Symp. Operating Syst. Des. Implement. (OSDI)}, 2018, pp. 409--425.
\bibitem{hoenig2001} J. M. Hoenig and D. M. Heisey, ``The abuse of power: The pervasive fallacy of power calculations for data analysis,'' \textit{Amer. Statistician}, vol. 55, pp. 19--24, 2001.
\bibitem{tlc} New York City Taxi and Limousine Commission, ``TLC trip record data,'' 2026. [Online]. Available: \url{https://www.nyc.gov/site/tlc/about/tlc-trip-record-data.page} (accessed Nov. 4, 2026).
\bibitem{bts} U.S. Bureau of Transportation Statistics, ``Airline on-time performance data,'' 2026. [Online]. Available: \url{https://www.transtats.bts.gov/} (accessed Nov. 4, 2026).
\bibitem{holm1979} S. Holm, ``A simple sequentially rejective multiple test procedure,'' \textit{Scand. J. Statist.}, vol. 6, no. 2, pp. 65--70, 1979.
\end{thebibliography}
\end{document}